\documentstyle[times,epsfig]{mn}
\input{epsf}

\newcommand{\bez}{\begin{eqnarray*}}
\newcommand{\eez}{\end{eqnarray*}}
\newcommand{\be}{\begin{equation}}
\newcommand{\ee}{\end{equation}}
\newcommand{\beq}{\begin{eqnarray}}
\newcommand{\eeq}{\end{eqnarray}}
\newcommand{\bc}{\begin{center}}
\newcommand{\ec}{\end{center}}

\newbox\grsign \setbox\grsign=\hbox{$>$} \newdimen\grdimen \grdimen=\ht\grsign
\newbox\simlessbox \newbox\simgreatbox \newbox\simpropbox
\setbox\simgreatbox=\hbox{\raise.5ex\hbox{$>$}\llap
     {\lower.5ex\hbox{$\sim$}}}\ht1=\grdimen\dp1=0pt
\setbox\simlessbox=\hbox{\raise.5ex\hbox{$<$}\llap
     {\lower.5ex\hbox{$\sim$}}}\ht2=\grdimen\dp2=0pt
\setbox\simpropbox=\hbox{\raise.5ex\hbox{$\propto$}\llap
     {\lower.5ex\hbox{$\sim$}}}\ht2=\grdimen\dp2=0pt

\def\simlt{\mathrel{\copy\simlessbox}}

\def\Tbb{T_{\rm bb}}
\def\Te{T_{e}}
\def\tT{\tau_{\rm T}}

\def\sigmaT{\sigma_{\rm T}}
\def\ldiss{l_{\rm diss}}
\def\Ldiss{L_{\rm diss}}
\def\Ls{L_s}

\def\Fh{F_h}
\def\Fbb{F_{\rm bb}}
\def\Astat{A_{\rm static}}
\def\Dstat{D_{\rm static}}
\def\TC{T_{\rm C}}

\def\mur{\mu_{\rm 0}}
\def\mus{\mu_{\rm s}}
\def\dd{{\rm d}}
\def\rme{{\rm e}}

\begin{document}

\title[X-ray spectra of dynamic coronae]
{X-ray spectra of accretion discs with dynamic coronae}

\author[J.~Malzac, A.~M.~Beloborodov and J.~Poutanen]
{\parbox[]{6.8in} {Julien~Malzac,$^{1,2\star}$
Andrei~M.~Beloborodov$^{3,4\star}$
and Juri~Poutanen$^{3\star}$}\\
$^1$Centre d'Etude Spatiale des Rayonnements (CNRS/UPS)
9, Av du Colonel Roche, 31028, Toulouse Cedex 4, France\\
$^2$Osservatorio Astronomico di Brera, via Brera, 28, 20121 Milano, Italy\\
$^3$Stockholm Observatory, SE-133 36 Saltsj\"obaden, Sweden \\
$^4$Astro Space Centre of Lebedev Physical
Institute, 84/32 Profsojuznaja Street, Moscow 117810, Russia}

\date{Accepted, Received}

\maketitle


\begin{abstract}
We compute the X-ray  spectra  produced  by {\it  non-static}  coronae  atop
accretion  discs around  black holes and neutron  stars.  The hot corona is
radiatively  coupled to the underlying disc (the reflector) and generates an
X-ray spectrum  which is sensitive  to the bulk  velocity of the coronal  
plasma, $\beta=v/c$. We  show  that an outflowing  corona  
 reproduces the hard-state spectrum of Cyg~X-1 and similar objects.  
The  dynamic  model  predicts a correlation  between the observed  amplitude  
of  reflection  $R$ and the X-ray spectrum slope  $\Gamma$  since both 
strongly  depend on $\beta$. A similar correlation was observed and its 
shape is well fitted by the dynamic model.
The scattering of soft radiation in an outflowing  corona can also account 
for the observed optical-UV polarization pattern in active galactic nuclei.
\end{abstract}

\begin{keywords}
{accretion, accretion discs --
radiative transfer -- gamma-rays: theory -- 
galaxies: Seyfert -- X-rays: general --
stars: individual: Cyg~X-1}
\end{keywords}


\section{Introduction}

The hard X-ray spectra of galactic  black holes (GBHs) and active  galactic
nuclei  (AGN)  indicate  the  presence  of  hot  plasmas  with  temperature
$k\Te\sim 100$~keV and scattering optical depth $\tT\sim 1$ in the vicinity
of accreting  black holes (see reviews by Zdziarski et al.  1997;  Poutanen
1998).  The  plasma  can be  identified  with  a  corona  of a  black  hole
accretion  disc (e.g.  Bisnovatyi-Kogan  \&  Blinnikov  1977;  Liang  1979;
Galeev, Rosner \& Vaiana 1979; see  Beloborodov  1999a for a recent review,
hereafter   B99a).  The   corona  is   likely   to  form  as  a  result  of
magnetorotational  instabilities  in  the  disc  and  the  buoyancy  of the
generated  magnetic  field  (Tout  \&  Pringle 1992; Miller \& Stone 2000). 
The corona is probably  heated in
flare-like  events of magnetic  dissipation  producing  the variable  X-ray
emission.

\footnotetext{$^\star$ E-mail: malzac@brera.mi.astro.it (JM);
andrei@astro.su.se (AMB); juri@astro.su.se (JP)}

The dominant  cooling  mechanism of the flaring plasma is Compton  cooling.
The observed hard X-rays are generated by the Comptonization  process, i.e.
by multiple  upscattering  of seed soft photons by the hot electrons in the
corona.  The Comptonization generally produces power-law X-ray spectra.  In
addition  to the  direct  power-law  radiation  from the  corona,  one also
observes  the  X-rays  reflected   by  the
underlying (relatively cold)  accretion  disc  (e.g.  George \&  Fabian  1991).  The  arising
reflection  features in the spectrum, in  particular,  the Compton bump and
the fluorescent iron line provide diagnostics for accretion models.

The    cold accretion disc partly  reemits the incident  X-rays in
the form of soft thermal  radiation.  This radiation   cools the corona,
providing the feedback loop which  regulates the  temperature of the corona
(Haardt \& Maraschi 1993, hereafter HM93).  
The  geometry of the corona can  hardly be
derived  from first  principles.  It might be a large  cloud  covering  the
whole  inner  region of the disc.  It may also be a number  of  small-scale
blobs with short life-times, with the energy release  concentrated in space
and time.  The resulting X-ray  spectrum is  however not sensitive to the
exact shape of the cloud, its density distribution, and other details.  The
only important parameter is the effective feedback factor (HM93; Stern et al.  
1995b; PS96; Svensson 1996) that is the fraction  of  the  X-ray luminosity   
which  reenters  the  source  after reprocessing.

Previous computations of the disc-corona models all assumed that the corona
is static  (e.g.  HM93; Stern et al.  1995b;  Poutanen  \&  Svensson  1996,
hereafter  PS96; see Svensson 1996, Poutanen 1998 for a review).  The model
was  successfully  applied  to  Seyfert~1  AGN,  however,  it was  found to
disagree with  observations of some  black-hole  sources in the hard state,
for  instance,   Cyg~X-1  (see  e.g.   Gierli\'nski   et  al.  1997  and
Section~\ref{sec:static}).  Three  alternatives  have been  suggested:  (i)
the cold disc is disrupted in the inner region (e.g.  Poutanen,  Krolik \&
Ryde 1997; Esin et al.  1998), (ii) the disc is highly ionised (Ross, Fabian
\& Young 1999;  Nayakshin,  Kazanas \& Kallman  2000), and (iii) the coronal
plasma is moving  away from the disc and emits  beamed  X-rays 
(Beloborodov   1999b, hereafter  B99b). We here study the third scenario.

B99b showed that the static model is not self-consistent for 
$e^{\pm}$-dominated flares and argued that the hot plasma should be 
ejected with a mildly  relativistic  bulk  velocity $\beta=v/c\la  0.5$. 
The bulk acceleration is also efficient in flares dominated by normal $e-p$ 
plasma if the compactness parameter of the flare exceeds $\sim 100$ (B99a).
The likely pattern of ejection resembles that of solar flares (scaled
to higher velocities because of higher compactness). The ejection
velocity may be directed  away or towards the disc (and it may change) with
the preferential  direction away from the disc.  

Mildly  relativistic  bulk
motion causes  aberration of the X-ray  emission and strongly  affects both
the amplitude of reflection $R$ and the spectrum slope  $\Gamma$.  Using
a  simple  analytical  model,  B99b  estimated  the  dependence  of $R$ and
$\Gamma$ on $\beta$ and found that the low amplitude of reflection  $R\sim
0.3$ and the spectrum  hardness  $\Gamma\sim  1.6$  observed in the hard
state of Cyg~X-1 can be  explained  with  $\beta\sim  0.3$.  The model also
explains the observed correlation  between $R$ and $\Gamma$
(B99a; Zdziarski,  Lubi\'nski \& Smith 1999; Gilfanov, Churazov \&
Revnivtsev 2000).

In the present  paper, we perform exact  computations  of the X-ray spectra
produced  by dynamic coronae.  We use a  non-linear  Monte-Carlo  code
(Malzac  \&  Jourdain  2000)  which is based on the  large-particle  method
(Stern  1985; Stern et al.  1995a).  The  formulation  of the  problem  and
details  of  simulations  are  described  in   Section~\ref{sec:setup}.  In
Section~\ref{sec:static},  we compute the spectra in the static case which
was well studied  previously and illuminate the problem of static models.
In  Section~\ref{sec:dynamic},   we  present  spectra  from  dynamic
coronae.    The    results    are    compared    with    observations    in
Section~\ref{sec:obser}.


\section{Setup}
\label{sec:setup}

\subsection{A heated cylinder}

\begin{figure}
\centerline{\epsfig{file=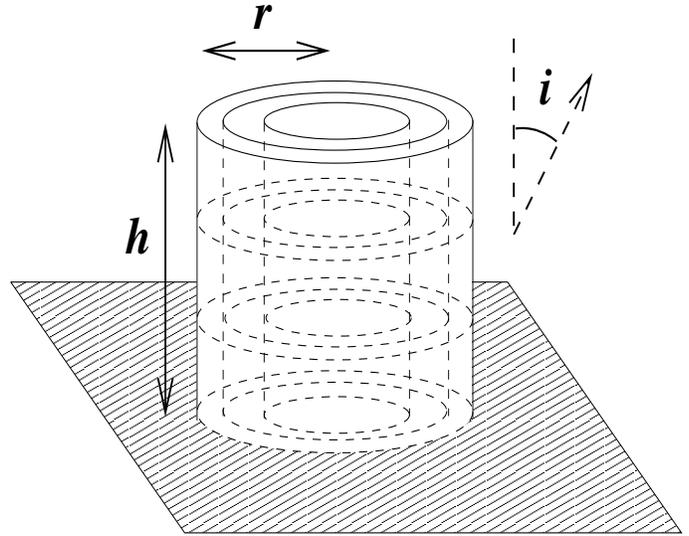,width=9.0cm,height=7.0cm}}
\caption{ A heated cylinder atop the accretion disc. 
The cylinder is divided into nine computational cells of equal volumes. 
The hot plasma moves through the cylinder with a velocity $\beta$.
}
\label{fig:geom}
\end{figure}

Consider a cylinder of radius $r$ and height $h$ located
atop  the  accretion  disc   (Fig.~\ref{fig:geom}).  The  cylinder  may  be
associated with a hot outflow  covering the disc or a heated  magnetic tube
in a compact  flare.  In the limit $r\gg h$ we get a slab  geometry  of the
corona,  which has the largest  feedback.  In the opposite  case, $r\ll h$,
the feedback  tends to zero (most of the  reprocessed  radiation  goes away
from the cylinder).  We study the sequence of geometries  parametrized with
the ratio $h/r$.  The plasma in the  cylinder is assumed to have a constant
density  and it is  heated  homogeneously  with a  constant  rate.

The plasma moves through the cylinder with a velocity $\beta$ directed 
normally to
the disc.  We assume that the heated cylindrical volume is static with 
respect to the disc.
The plasma  motion then implies a flux of particles  through the  cylinder.
If $\beta<0$  then the particle flux is absorbed by the disc at the bottom.
If $\beta>0$ then the particles  escape  through the top of the cylinder.
Outside the heated  region, the particles are  immediately  (on a
time-scale  $\ll  h/c$) cooled down to the Compton  temperature  of the
radiation field, $k\TC\sim  1-10$~keV.  In our  simulations, we do not take
into account  the scattering  on the escaped cold particles. 
Their density is likely to be  reduced  if the  plasma  
flows out along  diverging  magnetic lines.

In this paper, we restrict our consideration to thermal coronae and assume 
that the heated electrons have a  Maxwellian   distribution   with  a
temperature  $\Te$ in the plasma rest frame.  The  cylinder is divided into
nine cells of equal volumes (see  Fig.~\ref{fig:geom})  and the temperature
$\Te$ is calculated from the local  heating=cooling  balance in each of the
cells separately.  Note that the resulting  equilibrium  temperature is not
homogeneous.

\subsection{Feedback}

The X-rays from the corona strike the underlying disc and get  reprocessed.
We assume that reprocessed  radiation is the main cooler of the coronal
plasma and neglect soft radiation generated  viscously inside the accretion
disc.  This should be a good approximation if the corona releases energy in
the form of strong  concentrated  flares; then, locally, the soft flux from
inside   the   disc  is   much   smaller   than   the   reprocessed   
flux\footnote{Yet, the total surface-integrated intrinsic luminosity of the
cold disc  does not need to be much  smaller  than the  coronal  luminosity
since the covering  factor of the corona may be small (Haardt,  Maraschi \&
Ghisellini  1994).}.  We also  neglect  additional  photon  sources  such as
bremsstrahlung  and  cyclo-synchrotron  emission  in the corona  (see e.g.
Wardzi\'nski \& Zdziarski 2000).  An independent  argument in favour of the
reprocessed radiation as the main cooler comes from the observed $R-\Gamma$
correlation (Zdziarski et al.  1999; Gilfanov, Churazov \& Revnivtsev 1999;
see Section~\ref{sec:obser}).

We assume that the reflecting material of the disc is sufficiently dense so 
that the ionisation parameter $\xi\simlt 10^3$ and the ionisation effects are 
weak (e.g. \.Zycki et al. 1994). Then the albedo is small, $a\sim 0.2$, and
most of the X-rays  impinging  the disc are reprocessed.
We assume that the  reprocessed  flux 
has a quasi-blackbody spectrum (possibly diluted) 
with a constant temperature  $\Tbb$.  In  the  simulations,  we
consider  two cases:  $k\Tbb=150$~eV  and  $k\Tbb=5$~eV,  representing  the
typical blackbody temperatures in GBHs and AGN, respectively.

The  feedback  factor  $D$ is defined as the ratio of the soft  blackbody
luminosity  entering the cylinder, $\Ls$, to the total luminosity  outgoing
from the  cylinder,  $L$.  The  Comptonization  time-scale  is quite short,
$t_*\sim  (h/c)\log (m_ec^2 / 3k\Tbb)$, and we assume the heating rate to
be steady on  time-scales  $t\la t_*$.  Then the X-rays are  generated in a
quasi-stationary regime and  the  feedback  factor  determines  the  Compton
amplification factor of the hot plasma $A=L/L_s$ through the relation
\be \label{eq:DA}
 DA=1.
\ee
The amplification factor in its turn is the main magnitude  controlling the
spectral slope $\Gamma$ of the Comptonized radiation (see
B99a and Section~\ref{sec:gamma}).

\subsection{Radiative transfer and energy balance}

We  compute  the  plasma  temperature  simultaneously  with  the  radiative
transfer.  Our code is based on the non-linear  Large Particle  Monte-Carlo
method   (Stern  1985;  Stern  et   al.~1995a). In contrast to the standard
Monte-Carlo  technique,  this  method  allows  one to  follow  the path and
successive  interactions of photons and particles in parallel.  The code is
described and tested in Malzac \& Jourdain (2000).

We start a simulation  from an initial  (non-equilibrium)  state and follow
the evolution of the system plasma plus  radiation  until a steady state is
achieved.  The time step is about  $10^{-2}h/c$.  We compute the  evolution
of the  temperature  distribution  in the  cylinder  from the local  energy
balance.  The  plasma  gains  energy  owing to
constant  heating and loses  energy via Compton  cooling.  The  difference
between heating and cooling determines the change of the temperature during
each time step.

The radiative  transfer (photon  tracking) is dealt in the lab frame in the
same  way  as  for  static  coronae  (Malzac  \& Jourdain 2000).  The  only
difference  is that the velocity of the  scattering  electron is  generated
from an isotropic Maxwellian distribution in the plasma rest frame and then
it is Lorentz transformed to the lab frame.

In all simulations in this paper, we assume a constant  density  throughout
the  cylinder  and  parametrize   our  models  by  Thomson   optical  depth
$\tT=n_{e}\sigmaT h$ rather than   a dissipation  rate $L_{\rm diss}$
or   compactness   $\ldiss\equiv (\Ldiss/h)(\sigmaT/m_ec^3)$.   Strictly
speaking,  the model is  self-consistent  for low temperatures or 
low  compactnesses, when pair  production is not  important. In pair dominated 
blobs, the density cannot be assumed  homogeneous: the density  distribution
is then  determined  by the  local  pair  balance.  Yet,  models  with pair
production give similar results if they have same $\tT$.

The model has four parameters:  (i) Thomson optical depth $\tT$ (defined
along the  height of the  cylinder),  (ii)  height to  radius  ratio of the
cylinder  $h/r$, (iii) bulk velocity  $\beta$, and (iv) blackbody
temperature  $\Tbb$.  Given these  parameters the code computes the emitted
spectrum as a function of the inclination  angle $i$ (the angle between the
disc  normal and the line of sight).  We  consider  10 angular  bins of 
equal width $\Delta\cos i=0.1$.  The spectra are averaged within each bin.

\subsection{The intrinsic and reflected components}

For each inclination we compute two components of the observed emission: 

\begin{enumerate} 

\item
The `intrinsic' Comptonised X-rays, i.e. the photons coming to the observer 
directly after the upscattering in the hot blob. 
The slope of the intrinsic X-ray spectrum (the photon index $\Gamma$) is 
evaluated from a least-square  fit in the 2 -- 10 keV range.

\item
The reflected/reprocessed 
radiation from the underlying disc. We count only those reflected  photons 
that are  not  scattered  in  the blob, thus accounting for the 
attenuation of reflection by the blob.

\end{enumerate}

We model the  reflector  as an infinite  slab and  compute  the  reflection
component  assuming neutral  reflecting  material with standard  abundances
(Anders \& Ebihara  1982).  The amplitude of  reflection  is defined as the
ratio of an observed reflected component to that expected from an isotropic
point source  illuminating the slab,
\be\label{eq:defR}
R(i)=\frac{L_{\rm refl}(i)}{L_{\rm  refl}^{\rm  iso}(i)}.
\ee
The simulations yield the intrinsic  Comptonized  spectrum emitted
at an inclination $i$ and the reflected  luminosity $L_{\rm refl}(i)$.  
We then compute the
reflected  luminosity  $L^{\rm iso}_{\rm  refl}(i)$ for an isotropic point
source   with  the  same   intrinsic   spectrum  and  find   $R(i)$   from
equation~(\ref{eq:defR}).  
We  fitted  some of our  simulated  spectra  with the  {\sc  pexrav}  model
(Magdziarz  \& Zdziarski  1995) under  XSPEC.  The best fit values we found
for $R$ are very close to those derived using equation~(\ref{eq:defR}).


\section{Static coronae}

\label{sec:static}

\begin{figure*}
\centerline{\epsfig{file=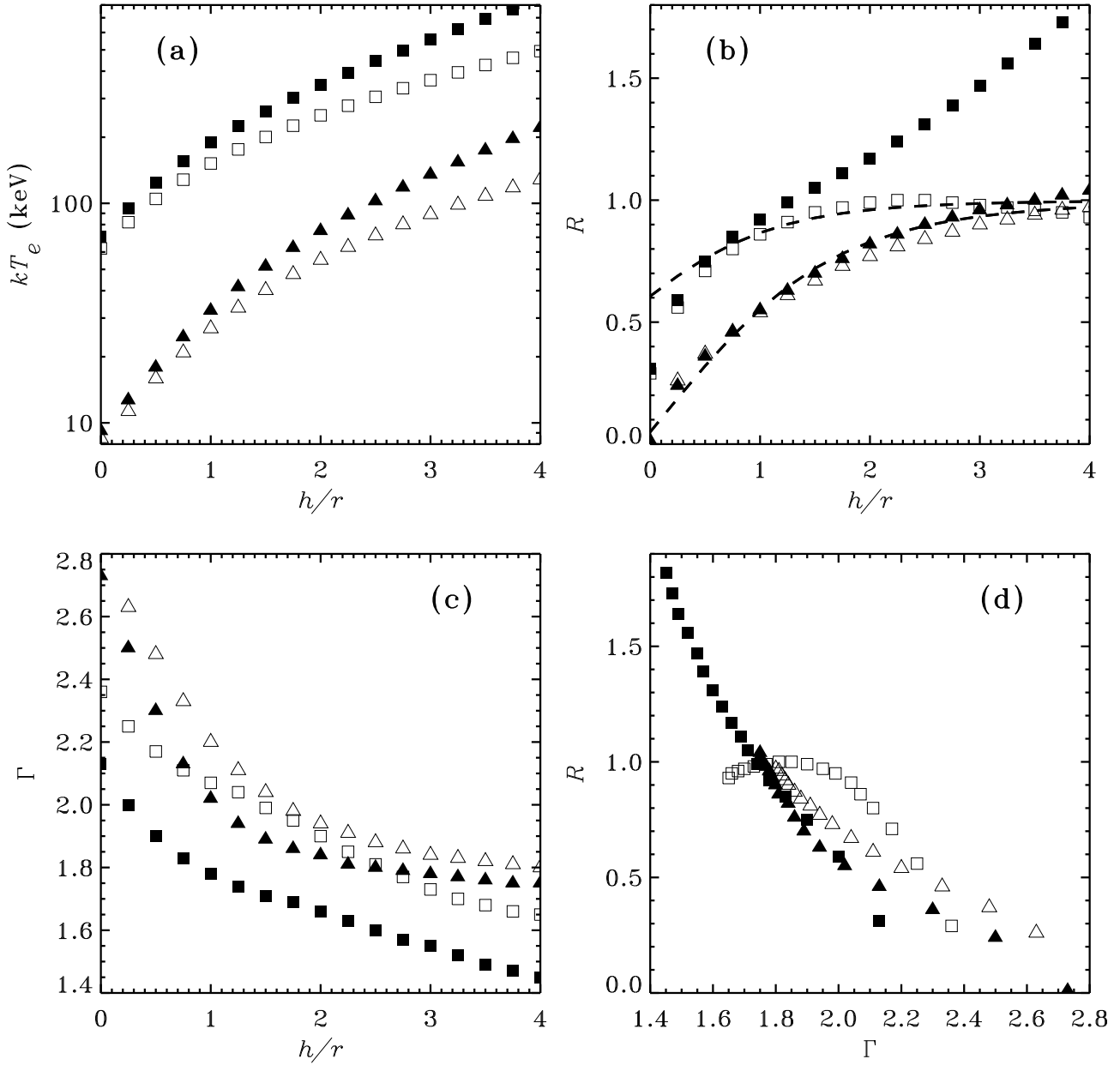,width=17cm,height=17cm}}
\caption{Characteristics of the static disc-corona model ($\beta=0$).
The active blob is a cylinder of height $h$ and radius $r$.
The model is determined by three parameters: $h/r$, the optical depth of the
cylinder, $\tT$ (measured in the vertical direction), and the temperature of
the soft radiation entering the blob, $\Tbb$.
Filled and open   symbols correspond to $k\Tbb=150$~eV (GBHs) and
$k\Tbb=5$~eV (AGN), respectively.
Squares and triangles show the cases $\tT=0.5$ and $\tT=3$.
(a) The average electron temperature in the blob versus $h/r$.
(b) The inclination-averaged amplitude of reflection $R$ versus $h/r$.
The dashed curves represent the analytical formula~(\ref{eq:refl0}).
(c) The inclination-averaged photon index of the intrinsic X-ray spectrum
$\Gamma$ (measured in the $2-10$ keV band) versus $h/r$.
(d) $R$ versus $\Gamma$.
} \label{fig:static}
\end{figure*}

We have computed a set of static models ($\beta=0$) with  different  $h/r$,  
$\tT$, and $\Tbb$, and  determined $R$ and $\Gamma$ in the  calculated  
spectra. The results  are  shown in  Fig.~\ref{fig:static}.  
For  simplicity,  the presented amplitude of reflection is averaged over
inclinations, $R=\int_0^1 R(i) \dd (\cos i)$.

\subsection{Feedback, amplification, and the spectral index}

\label{sec:feedstat}

\begin{figure*}
\centerline{\epsfig{file=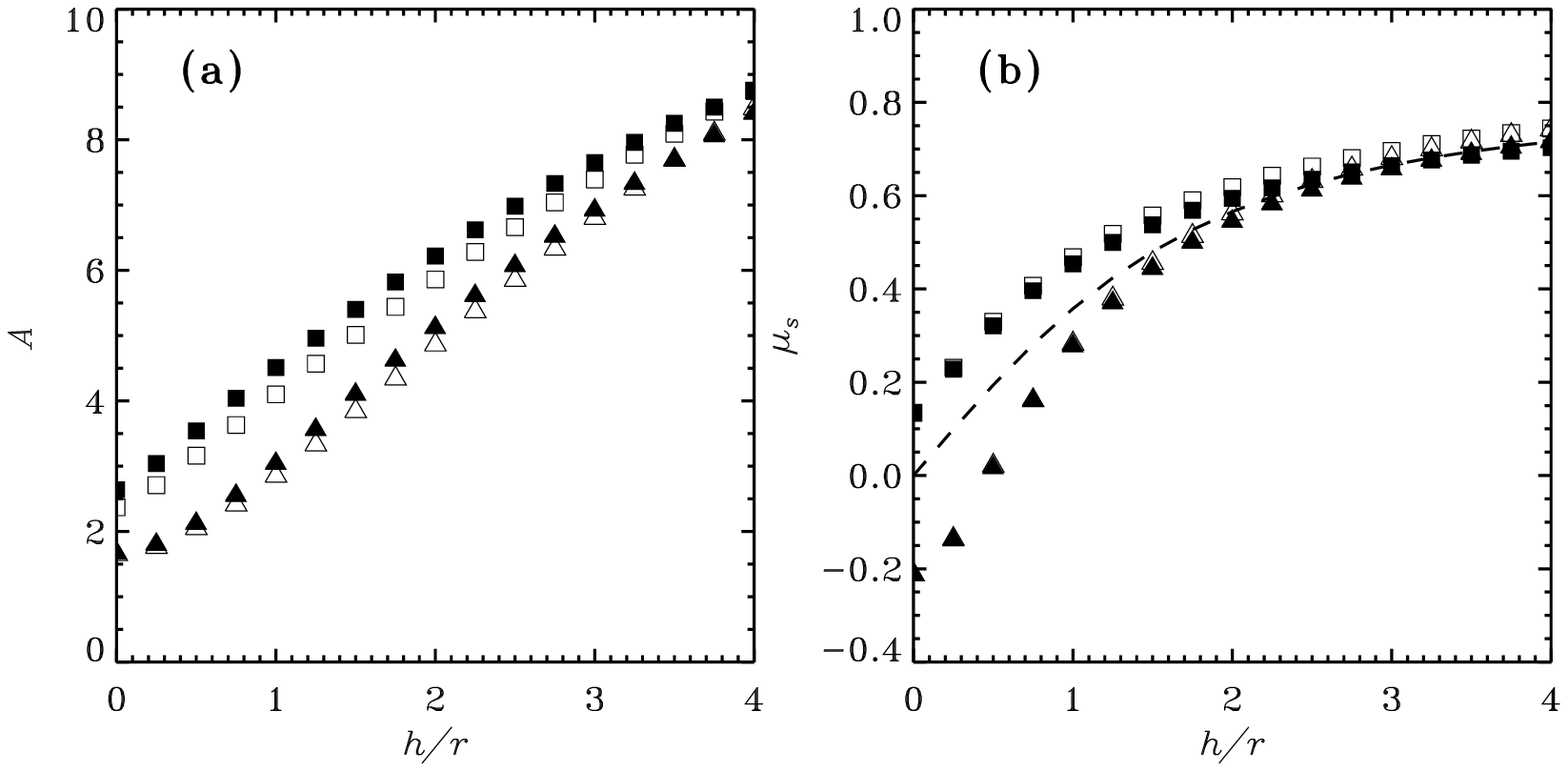,width=17cm,height=8cm}}
\caption{
(a) Amplification factor, $A$, and (b) the effective geometrical factor $\mus$
(see eq.~\ref{eq:mus}) for static coronae.
The symbols have same meaning as in Fig.~\ref{fig:static}.
The dashed curve shows $(4/5)\mur$, the $\mur$ is given in 
equation~(\ref{eq:refl0}). 
} \label{fig:amplmus}
\end{figure*}

In the static case,  the feedback factor  $\Dstat$ is determined
mainly by the geometry of the cloud, i.e.  by $h/r$ in our parametrization.
We express the feedback factor as
\be\label{eq:mus}
  \Dstat=\Astat^{-1}=\frac{1}{2}(1-a)(1-\mus),
\ee
where $a$ is the energy-integrated albedo of the disc. 
This equation defines the
effective  geometrical  parameter  $\mus$ for a static cloud (B99a,b).  The
difference  $1-\mus$  is the  fraction  of  reprocessed 
radiation  that  returns to the hot cloud.
To the first approximation,   $\mus$  can  be  associated  with
$\mur\equiv  h/(4r^2+h^2)^{1/2}$,   where $1-\mur$ corresponds to the 
angular size of
the  cylinder  base  measured   from  its  centre.  Numerically,   we  find
$\mus\approx (4/5)\mur$ (see Fig.~\ref{fig:amplmus}b).

\begin{figure*}
\centerline{\epsfig{file=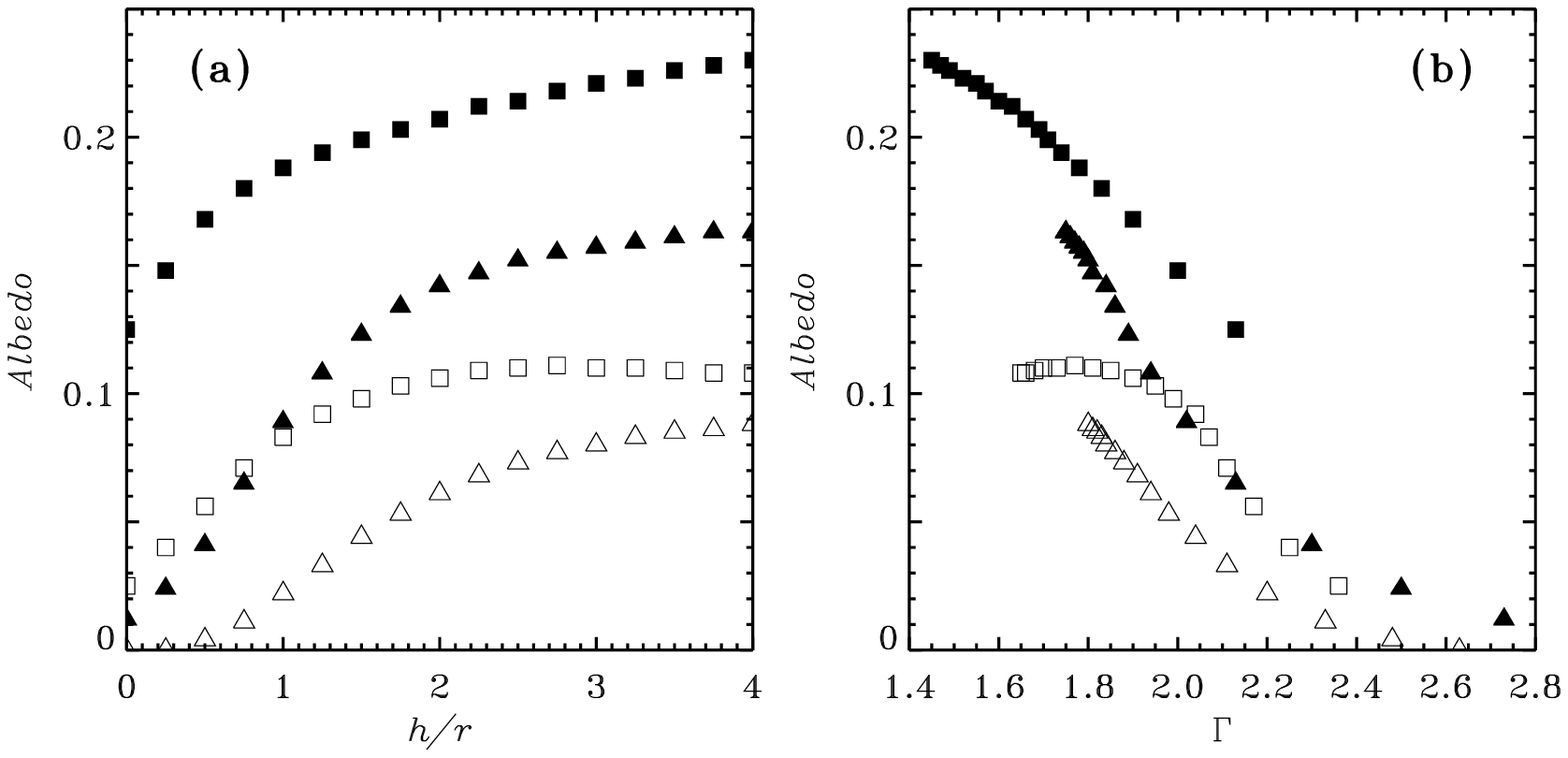,width=17cm,height=8cm}}
\caption{
Energy-integrated albedo, $a$, as a function of (a) $h/r$ and (b) spectral index
$\Gamma$. The symbols have same meaning as in Fig.~\ref{fig:static}.
} \label{fig:albedo}
\end{figure*}

As one can see from  Fig.~{\ref{fig:amplmus}}, the effective $\mus$ is not fully
determined by $h/r$ but also depends on $\tT$.  The strongest  dependence on
$\tT$ is observed for flat geometries ($h/r\la 1$):
$\mus$ and $A$ substantially decrease  as  $\tT$  increases.  This  effect  can be
understood  analytically.  Consider a slab corona and denote the  radiation
fluxes outgoing from the slab at the bottom and the top as $F_-$ and $F_+$,
respectively.  The  reprocessed  flux  entering  the  slab  from  below  is
$F_s=(1-a)F_-$.  The amplification factor is, by definition,
\be 
A=\frac{F_-+F_+}{F_s}=\frac{1+F_+/F_-}{1-a}.
\ee  
One can estimate the ratio $F_+/F_-$ by solving the
radiative  transfer equation in the Eddington approximation 
(see Rybicki \& Lightman 1979, ch.~1) and assuming isotropic scattering.
Taking into
account that the energy sources are homogeneously  distributed in the slab,
we  get   
\be 
\frac{F_+}{F_-}=\frac{1+2/\sqrt{3}}{1+\sqrt{3}\tT/2}, \qquad
\mus=\frac{4-3\tT}{4(1+\sqrt{3})+3\tT}. 
\ee  
These expressions are in good agreement with the results 
of simulations (see Fig.~\ref{fig:amplmus}, $h/r=0$). 
If $\tT$ is large,
the radiation is trapped at the bottom, near the reflector.
 In the limiting case  $\tT\gg 1$, one has $F_+\ll  F_-$,  $A\rightarrow
(1-a)^{-1}$  and  $\mus\rightarrow  -1$.  
Then almost all
the X-rays are  reprocessed  because the source  emits most of them towards
the  reflector.  The bottom layers of the slab are cold and emit a
soft  spectrum  for  which  the disc  albedo  $a$ is  close  to zero  (see
Fig.~\ref{fig:albedo}b) and $A\rightarrow 1$.  Such a small $A$ results in
the high $\Gamma$ (see Fig.~\ref{fig:static}c).

At small $\tT$, the emitted spectrum hardens (Fig.~\ref{fig:static}c).
This is caused by the fact that 
the number of photons  scattered  in an  optically  thin blob is small. 
When the
Comptonized  spectrum is made of fewer photons, the same luminosity can
be emitted  only if the  average  photon  energy is high.  This  leads to a
hard  spectrum  (Fig.~\ref{fig:static}c)  with a break at high energies.
The optically thin models have high temperatures (Fig.~\ref{fig:static}a):
at fixed  geometry  and $\Tbb$ the product  $\tT T_e$ stays  approximately
constant   (HM93;   Stern   et  al.  1995b;   Svensson   1996;   see   also
Fig.~\ref{fig:tetau}) and hence an  optically  thin blob must be hotter.

Blobs with high $h/r$  (`elongated  cylinders') have low feedback and the
resulting  spectrum is hard  (Fig.~\ref{fig:static}c).  This  behaviour  is
known  from  previous  simulations  (see  Svensson  1996 for a review;  the
previously  computed spectra of detached   spheres are similar to those
of  elongated  cylinders).  The  product  $\tT T_e$ increases  with
$h/r$;  for fixed $\tT$ it implies a growth of the plasma temperature with
$h/r$ (Fig.~\ref{fig:static}a).

The  emitted  spectrum  depends  on  $\Tbb$.  For this  reason,  we perform
computations for GBHs  ($k\Tbb=150$~eV) and AGN ($k\Tbb=5$~eV)  separately.
For the same $h/r$ and  $\tT$, we get a harder  spectrum  for GBHs than for
AGN  (despite  the fact  that  the  feedback  factor  and $A$ are not   
sensitive  to  $\Tbb$).  The hardening is caused by two reasons:
(i) The  energy  range in which the
Comptonized  luminosity  is liberated  is narrower in GBHs.  Therefore,  in
order to radiate a given amount of  luminosity,  the spectrum has to harden
and cut off at higher energy.
(ii) Low-$\tT$ models are hot and have an anisotropy break in the spectrum
(Stern et al. 1995b). In GBHs, this break is in the 2-10 keV band and affects
the measured~$\Gamma$.

\begin{figure*}
\centerline{\epsfig{file=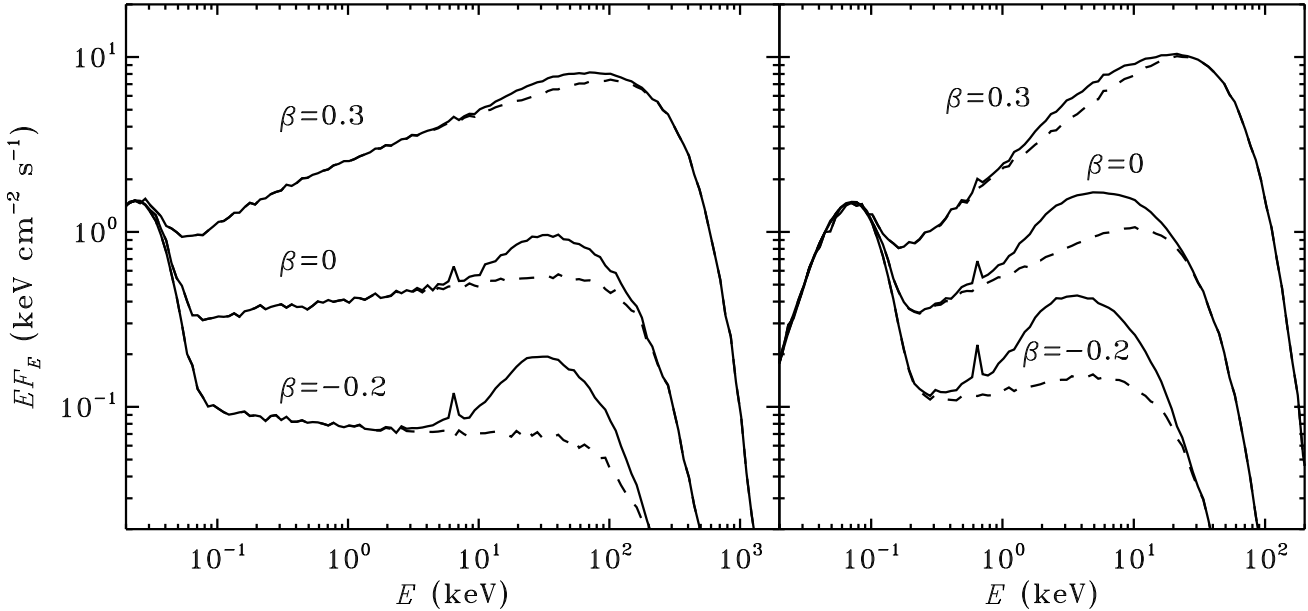,width=17cm,height=8cm}}
\caption{The effect  of bulk  motion  on the  emitted  spectra. Here
$h/r=2$, $\tT=3$, and nearly face-on inclination is assumed, 
$0.9<\cos i<1$. The spectra are normalised to the blackbody peak.
Left panel shows the case of AGN ($k\Tbb=5$ eV) and
right panel  corresponds to GBHs ($k\Tbb=150$ eV).
}
\label{fig:threespec}
\end{figure*}


\subsection{The amplitude of reflection}
\label{sec:statrefl}

The reflection  component produced beneath the blob is partly scattered and
attenuated when passing through the blob. To the first approximation,
the attenuation effect can be described analytically,
\be\label{eq:refl0}
   R=\mur+(1-\mur) \rme^{-\tT(1-\mur)}, \qquad
   \mur=\frac{h/2}{\sqrt{r^2+(h/2)^2}}.
\ee
Here it is assumed that the fraction  $\mur$ of the  reflected  luminosity
goes  directly  to the  observer.  The  remaining  fraction,  $1-\mur$,  is
reflected  at the base of the  cylinder  and  attenuated  when  transmitted
through the hot plasma.  The cylinder  optical  depth seen by soft  photons
emitted at the base  depends on $\tT$ and $h/r$; it equals  $\sim  \tT$ for
$h/r=0$  and  vanishes  at  $h/r\gg  1$.  We  interpolate  between  the two
extremes as  $\tT(1-\mur)$  and put this  effective  optical depth into the
exponential in equation~(\ref{eq:refl0}).

The attenuation formula~(\ref{eq:refl0}) assumes that 
the scattered reflection component is completely destroyed. Indeed, the 
`Comptonised reflection' component has a power-law spectrum where the 
reflection features (the Fe lines and the edge) disappear.

Equation~(\ref{eq:refl0})  is in good  agreement  with the  simulation
results (see  Fig.~\ref{fig:static}b).  The attenuation effect is strong if
$h/r$ is small (so that a large fraction of the reflected  radiation has to
pass through the cylinder) and if $\tT$ is large.  In the  optically  thick
regime, $\tT\ga 3$, almost all reflected  photons  reentering the cloud are
Compton scattered and the attenuation  effect  saturates:  further increase
in $\tT$ does not reduce $R$.

For elongated  cylinders, the attenuation  affects only a small fraction of
the   reflected   luminosity and $R$   approaches   unity   at  high  $h/r$. 
The reflection  amplitude is higher in GBHs,  especially in the
high-temperature  models  (with small $\tT$ and large $h/r$) where $R$ 
substantially exceeds unity (see Fig.~\ref{fig:static}b).  
The  explanation  is as  follows. The radiation scattered once 
is strongly  anisotropic  and   collimated 
towards the reflector (HM93). In GBHs, because of relatively 
large temperature of seed soft photons, the first  scattering order 
peaks above 1~keV and contributes significantly to the reflected
flux.  As a result, $R$  increases.  By contrast,  in AGN, the  Comptonized
emission  above 1 keV is  made  of  multiply  scattered  photons  and  the
anisotropy effects are small.


\subsection{$R-\Gamma$ anticorrelation}

Models with small $h/r$ have soft spectra because of large feedback.  These
models also have lower reflection amplitudes
because of stronger  attenuation   (see  Section~\ref{sec:statrefl}).
Therefore,
variations in  $h/r$  produce  an  anticorrelation  between  $R$  and  $\Gamma$
(Fig.~\ref{fig:static}d).  
The  anticorrelation  disagrees  with  the data
(see  Section~\ref{sec:obser}).  A static  model  with a neutral  reflector
never predicts a small $R$ simultaneously with a hard spectrum.  
In particular,
the  parameters  $R\sim 0.3$ and  $\Gamma\sim  1.6$ observed in Cyg~X-1 and
similar hard-state objects cannot be reproduced.


\section{Dynamic coronae}

\label{sec:dynamic}

\subsection{Example spectra}

In the  dynamic  case,  $\beta\neq  0$, the  corona  spectrum  is  strongly
affected  by the bulk  motion of the hot  plasma.  This is  illustrated  in
Fig.~\ref{fig:threespec}   where  we  compare  the  cases  $\beta=0.3$  and
$\beta=-0.2$  with the static case.  The spectra are computed  for the same
optical depth, $\tT=3$, and geometry, $h/r=2$.

In the case of $\beta=0.3$  (plasma moves away from the disc), the observed
Comptonized luminosity is enhanced as a result of relativistic  aberration.
The X-rays are beamed away from the disc, and the reprocessed and reflected
luminosities  are  reduced.  The low  feedback  leads  to a hard  intrinsic
spectrum.

In  the  case  of  $\beta=-0.2$   (plasma  moves  towards  the  disc),  the
Comptonized  luminosity is beamed towards the disc and the  reprocessed and
reflected  components  are  enhanced.  The high  feedback  leads  to a soft
intrinsic spectrum.

\begin{figure}
\centerline{\epsfig{file=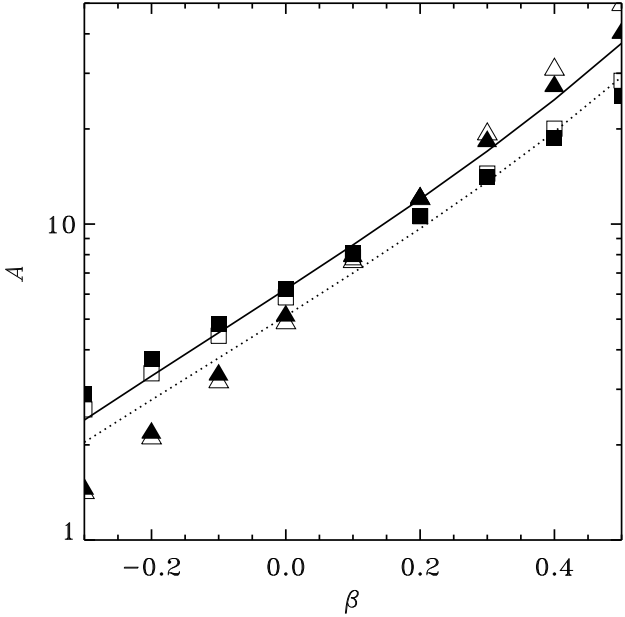,width=8.5cm,height=8.5cm}}
\caption{Amplification factor $A$ as a function of $\beta$.
The simulations are done for a cylinder with $h/r=2$. The results are shown
by symbols with the same meanings as in Fig.~\ref{fig:static}.
The solid and dotted curves display the analytical formula (\ref{eq:A})
for GBH models with $\tT=0.5$ and $\tT=3$, respectively.
}
\label{fig:amplbeta}
\end{figure}

Since  $\tT$ is fixed in  Fig.~\ref{fig:threespec},  a high (low)  feedback
leads to a low  (high)  coronal  temperature.  This  causes  the shift of the
spectral  break to lower  energies with  decreasing  $\beta$.  

The moderate
changes in the amplitude  and  direction of the bulk  velocity  thus induce
crucial  changes  in  the  emitted  spectra.  Note  that  in  most  of  our
simulations,  the bulk  velocity  is  smaller  than  the  thermal  electron
velocity   $\beta_{\rm   th}$   (e.g.   $\beta_{\rm   th}\sim   0.5$   for
$k\Te=100$~keV).

\begin{figure}
\centerline{\epsfig{file=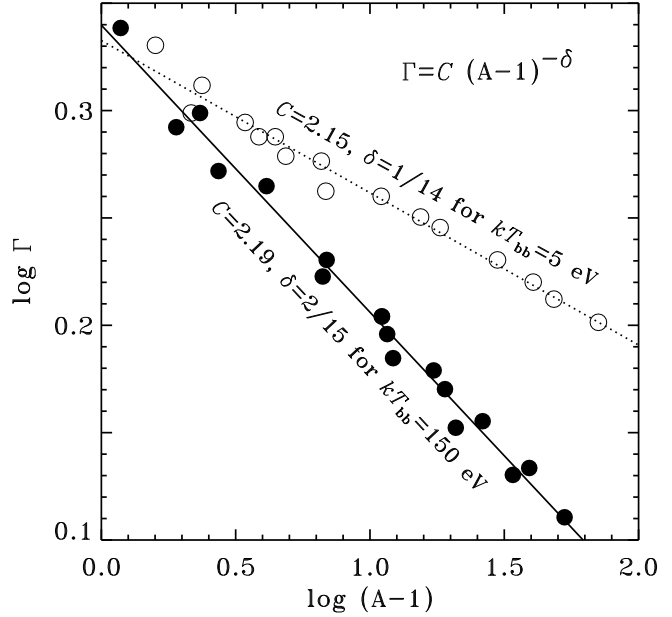,width=8.5cm,height=8.5cm}}
\caption{
$\Gamma$ versus $A$. The results of simulations for GBHs and AGN
are shown by filled and open circles, respectively.
The source geometry is fixed (a cylinder with $h/r=2$), while $\tT$ and
$\beta$ vary. Only models with $50<kT_e<300$~keV are shown here.
The lines correspond to approximation~(\ref{eq:gam_ampl}).
}\label{fig:gam_ampl}
\end{figure}


\subsection{The amplification factor}

The low feedback at $\beta>0$ implies a high Compton  amplification  factor
of the X-ray source  (eq.~\ref{eq:DA}).  The main  effect can be estimated
analytically assuming that the source is isotropic in the plasma rest frame
(see eq.~7 in B99b),
\be \label{eq:A}
A(\beta)=\Astat
\frac{\gamma^2(1+\beta)^2(1+\beta\mus)^2}{1-\beta^2(1+\mus)^2/4}.
\ee
Here, $\gamma=1/\sqrt{1-\beta^2}$
and  $\mus$ is the effective geometrical parameter which we take from the
static model (see eq.~\ref{eq:mus} and Fig.~\ref{fig:amplmus}b); a more exact 
definition of $\mus$ is given in B99b. $A=\Astat$ at $\beta=0$ and 
monotonically increases with $\beta$; at $\beta<0$  equation~(\ref{eq:A}) gives 
$A<\Astat$.

In  Fig.~\ref{fig:amplbeta}  we compare the results of simulations with the
analytical  estimate~(\ref{eq:A}).  
In the optically thin models ($\tT=0.5$), we observe
the   reduction   in  $A$  for   large   positive   $\beta$   compared   to
equation~(\ref{eq:A}).  This is caused by the anisotropy of scattering in the
plasma rest frame  (optically thin models have very high  temperature,  see
Fig.~\ref{fig:moving}a, and scatter preferentially downwards). 
At $\beta<0$, $\Te$ is small and the agreement is good.

In the optically  thick models  ($\tT=3$),  we see that the actual value of
$A$ at $\beta<0$ is smaller than that given by equation~(\ref{eq:A}).  This
is caused by the trapping of radiation and its advection downwards by the
moving plasma. The advection enhances the anisotropy of the blob radiation  
compared to the optically thin case. 
In the limiting case of large  optical  depths, the velocity
of photon diffusion  $\beta_{\rm  diff}\approx  1/\tT$ is smaller than
the  advection  velocity,  $\beta$, so that almost no radiation  can escape
through the upper boundary.  When viewed from the 
plasma rest frame, the effect can be understood by noting that both the bottom 
and top boundaries of the cylinder move upwards;
as a result, the escape probability is larger for photons
emitted towards the bottom boundary.

For the similar reason,  equation~(\ref{eq:A})
underestimates  $A$ for high-$\tT$ models at $\beta>0$.  
In this case, the  radiation is advected
towards the upper  boundary, and it is strongly  beamed away from the disc,
more than in the optically thin case.

\subsection{The spectral index}
\label{sec:gamma}

A simple functional shape  $\Gamma(A)$  suggested in B99a fits the results of
the simulations,
\be\label{eq:gam_ampl}
\Gamma=C (A-1)^{-\delta}.
\ee
The fitting  parameters are $C=2.19$,  $\delta=2/15$  for GBHs ($k\Tbb=150$
eV) and $C=2.15$,  $\delta=1/14$  for AGN ($k\Tbb=5$ eV).  These parameters
are not far from  those  found by B99a in the case of a spherical blob:
$C=2.33$  for both GBHs and AGN and  $\delta=1/6$  and $1/10$  for GBHs and
AGN, respectively.  Equation~({\ref{eq:gam_ampl}})  is a good approximation
for  models  with  $k\Te$  in  the  range   between  50  keV  and  300  keV
(Fig.~{\ref{fig:gam_ampl}}).

\begin{figure*}
\centerline{\epsfig{file=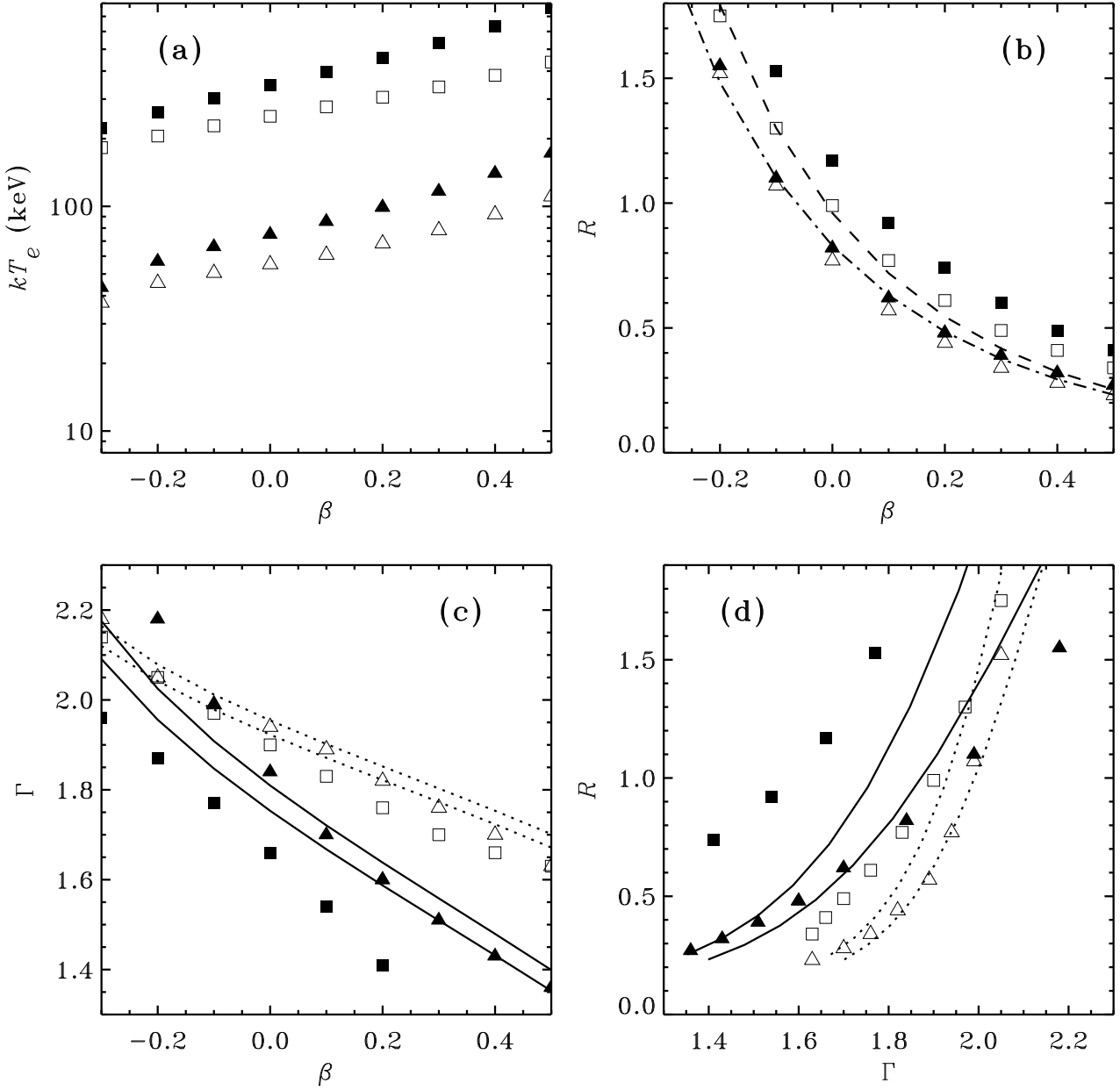,width=17cm,height=17cm}}
\caption{
Characteristics of the dynamic  disc-corona model.  The model is determined
by four  parameters:  $h/r=2$,  the optical  depth of the  cylinder,  $\tT$
(measured in the vertical direction), the temperature of the soft radiation
entering the blob,  $\Tbb$, and the  velocity of bulk motion  $\beta$.  The
observed characteristics are inclination-averaged. Filled and open symbols
correspond to $k\Tbb=150$~eV  (GBHs) and $k\Tbb=5$~eV (AGN), respectively.
Squares and triangles show the cases  $\tT=0.5$ and $\tT=3$,  respectively.
(a) The average  temperature in the blob versus $\beta$.
(b) The amplitude of  reflection  $R$  versus $\beta$.
The curves give analytical $\mu$-averaged $R$ from
equation (\ref{eq:reflbulk}) (dashed -- $\tT=0.5$, dot-dashed -- $\tT=3$).
(c) The photon  index of the intrinsic X-ray spectrum  
$\Gamma$  (in the 2-10 keV band)  versus  $\beta$.
The analytical approximation (eq. \ref{eq:A} and \ref{eq:gam_ampl}) is shown
by solid curves for GBHs and dotted curves for AGN.
(d) Reflection $R$ versus spectral slope $\Gamma$. The analytical model 
is shown by solid curves for GBHs and dotted curves for AGN (left curves for
$\tT=0.5$ and right curves for $\tT=3$).
}\label{fig:moving}
\end{figure*}

Equation~(\ref{eq:gam_ampl}) combined with equation~(\ref{eq:A}) yields
the analytical dependence $\Gamma(\beta)$ which we find to be in good
agreement  with the results of simulations (see Fig.~\ref{fig:moving}c).
Only for GBHs models with very high electron
temperatures ($k\Te\ga 300$~keV) the deviations
$\Delta\Gamma$ exceed 0.1. The reasons of these deviations are the same as 
in the static case discussed   in the end of Section~3.1.

\begin{figure*}
\centerline{\epsfig{file=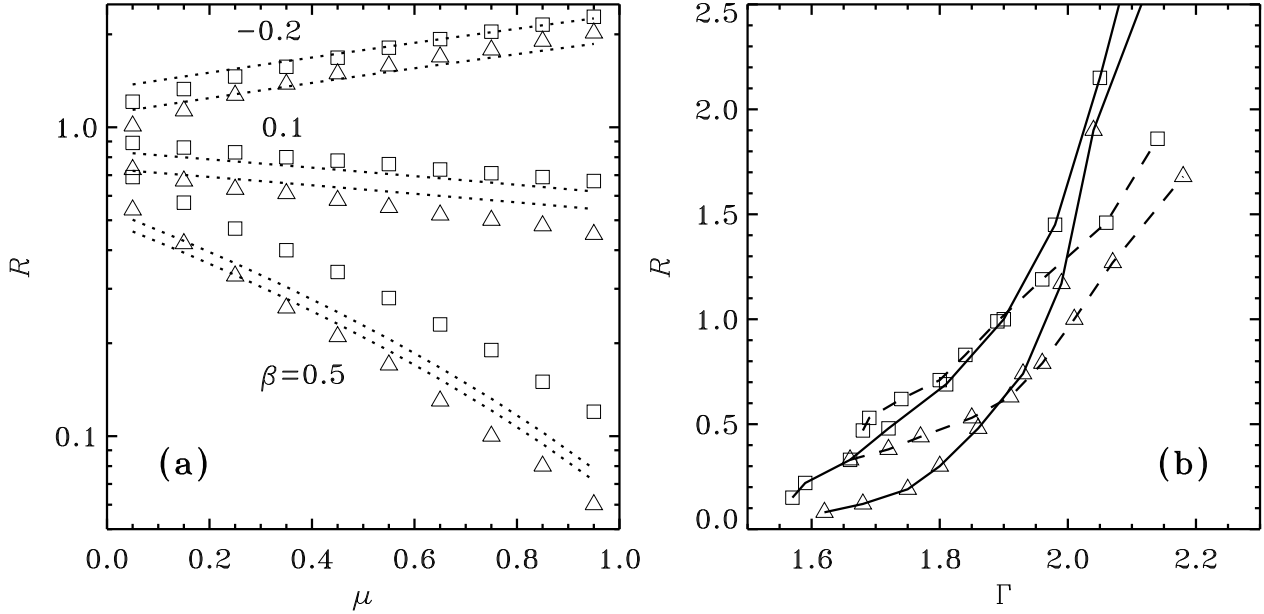,width=17cm,height=8cm}}
\caption{(a) Reflection,  $R$, versus  inclination  ($\mu=\cos i$).
Squares are for
$\tT=0.5$,   triangles  for  $\tT=3$.  Here  $h/r=2$,   $k\Tbb=5$  eV,  and
$\beta=-0.2,  0.1,  0.5$.  Reflection  is larger  that  unity for  negative
$\beta$ (bulk motion directed  towards the reflector),  and is much smaller
than  unity  for  the  outflows.  The  dotted  curves  are  the  analytical
estimates given by equation~(\ref{eq:reflbulk}).
(b) $R-\Gamma$  correlation  for
two inclinations  $\mu=0.25$  ($i=76\degr$, dashed curves), and $\mu=0.85$
($i=32\degr$,  solid  curves).  Here  $\beta$  varies from $-0.3$ 
(upper points) to $0.5$ (lower points) with the step 0.1. 
The curves connecting the points are added for clarity.
}
\label{fig:rvsmu}
\end{figure*}

\subsection{The amplitude of reflection}

Following the analysis of B99a,b, but accounting for the attenuation effect 
(see Section~\ref{sec:statrefl}), we get an analytical formula for the 
reflection amplitude,
\beq
\label{eq:reflbulk}
R(\mu)&=&\frac{(1-\beta\mu)^3}{(1+\beta\mur)^2}
  \left\{\mur\left(1+\frac{\beta\mur}{2}\right) \right. \\
 &+&\left. \frac{(1-\mur)\left[1+\beta(1+\mur)/2\right]}{(1+\beta)^2}\;
  \rme^{-\tT(1-\mur)} \right\}, \nonumber
\eeq
where, $\mu=\cos i$, and $i$ is the disc  inclination.  Here, the reflected
luminosity  is  represented  as a sum  of  two  parts:  the  first  one  is
reflected outside the cylinder base and does not experience any attenuation
and the second one is reflected from the base and it is partially  attenuated,
depending on $\tT$.  When $\mur$ approaches unity (or $\tT\rightarrow  0$),
the attenuation is not important and  equation~(\ref{eq:reflbulk})  becomes
equation~(3) in B99b:
\be \label{eq:reflappx}
R(\mu)=\frac{(1+\beta/2)(1-\beta\mu)^3}{(1+\beta)^2}.
\ee
Equation~(\ref{eq:reflbulk}) is in good agreement with the simulations (see
Fig.~\ref{fig:moving}b).  Substantial  differences  appear only in the case
of  low-$\tT$  GBH  models.  The  differences  are  caused  mainly  by  the
anisotropy of scattering in the rest frame of the hot plasma.  The similar
effect was observed in the static GBH models (Fig.~\ref{fig:static}b) and 
discussed in Section~\ref{sec:statrefl}.

The left panel of  Fig.~\ref{fig:rvsmu}  shows the dependence of reflection
on the disc  inclination.  Note that $R$ is less  sensitive  to  $\beta$ at
large  inclinations.  The overall  angular  dependence of $R$ is quite well
represented by equation~(\ref{eq:reflbulk}).

\subsection{$R-\Gamma$ correlation}

Fig.~\ref{fig:moving}d shows $R$ versus $\Gamma$. 
Again we see that the analytical formulae are in quite good agreement 
with the results of simulations.  
The $R-\Gamma$ correlation is steeper in AGN than in GBHs.
This is caused by the fact that $\Gamma(A)$ is a flat function at
small $\Tbb$ (see Section 4.3; the index $\delta$ in 
eq.~\ref{eq:gam_ampl} is smaller for AGN than that for GBHs).

The shape of the  $R-\Gamma$  correlation  also  depends on the disc
inclination: 
the reflection amplitude is very sensitive to $\beta$ at small inclinations
and the correlation gets stretched in the vertical direction 
(see Fig.~\ref{fig:rvsmu}b).

\subsection{Pair dominated coronae}

A compact and energetic blob emitting hard X-rays may 
be mainly composed of $e^{\pm}$ pairs produced in 
$\gamma-\gamma$ reactions. In this case, the blob optical depth and temperature
are determined by
the compactness parameter $\ldiss=\Ldiss\sigmaT/hm_ec^3$, where $\Ldiss$
is the power dissipated in the blob (see e.g. Stern et al. 1995a,b). The
equilibrium $\tT$ is controlled by the balance between pair
production and annihilation and $T_e$ is controlled by the energy balance.  

\begin{figure} 
\centerline{\epsfig{file=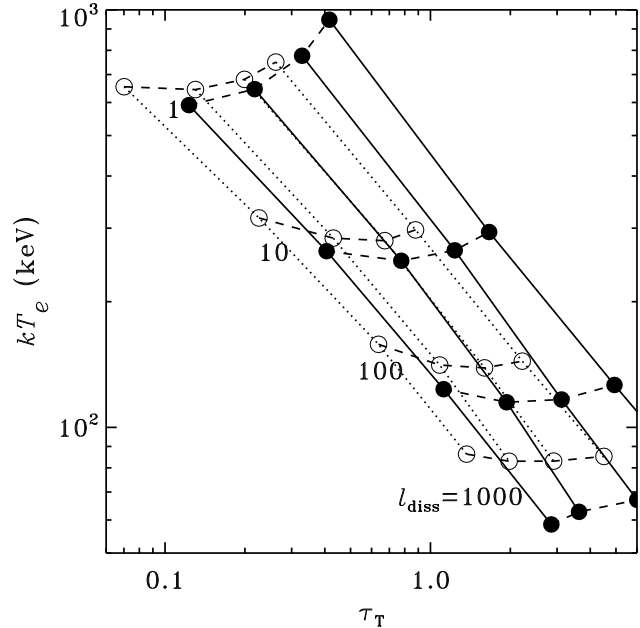,width=8.5cm,height=8.5cm}}
\caption{
The relation between $\Te$ and $\tT$ for $\beta=-0.3, 0.0, 0.3, 0.5$ (from
left to right). The simulations are done for a cylinder with $h/r=2$;  
the open and filled circles show the AGN and GBH models, respectively.
The $T_e-\tT$ relation is shown by dotted curves for AGN and solid curves for 
GBHs.  This relation  does not  depend  on the  nature  of the
scattering plasma. The dashed curves, describing the case of pure $e^\pm$
plasma,  connect  points  of equal dissipation compactnesses  $\ldiss$.
}
\label{fig:tetau}
\end{figure}

We have calculated a set of pair-dominated dynamic models with the $h/r=2$
geometry. For simplification, we compute a global pair balance assuming
a uniform pair distribution in the cylinder.  
The results are shown in Fig.~\ref{fig:tetau}. 

At fixed $\beta$, the model follows an equilibrium $\Te-\tT$ curve.
We emphasize that the $\Te-\tT$  relation for $e^{\pm}$ plasma is the 
same as for normal $e-p$ plasma,  since  it is determined by 
the energy balance only (see HM93; Stern et al.  1995b; PS96) and the 
annihilation radiation does not contribute much to the total energy budget. 

With increasing $\beta$, the $\Te-\tT$ curve moves  to the right.
An increase in  $\beta$ at fixed $\ldiss$ leads to an increase in 
the $e^\pm$ optical depth while the  temperature  does not change much.
This is the `thermostat' effect of pairs discussed previously in the
static models (see Figs~2 and 3 in Svensson 1996).
The  typical values $k\Te\sim100$ keV and $\tT\sim1-2$ inferred from 
observations (Zdziarski et al.  1997; Poutanen 1998) are obtained for large 
but possible values of the dissipation  compactness, $\ldiss>100$.  
 
We now briefly discuss the $R-\Gamma$  correlation expected at fixed 
$\ldiss$ and varying $\beta$. The correlation will be steeper compared
to the $\tT={\rm const}$ case shown in Fig.~\ref{fig:moving}d.
At large $\beta$, the equilibrium $\tT$ is large and here the 
$R-\Gamma$ track is close to that with $\tT=3$.
At $\beta<0$, $\tT$ drops and the $R-\Gamma$ track approaches the
$\tT=0.5$ curve.


\section{Comparison to Observations}
 \label{sec:obser}

\subsection{Broad-band spectra of individual sources}

\begin{figure*} 
\centerline{\epsfig{file=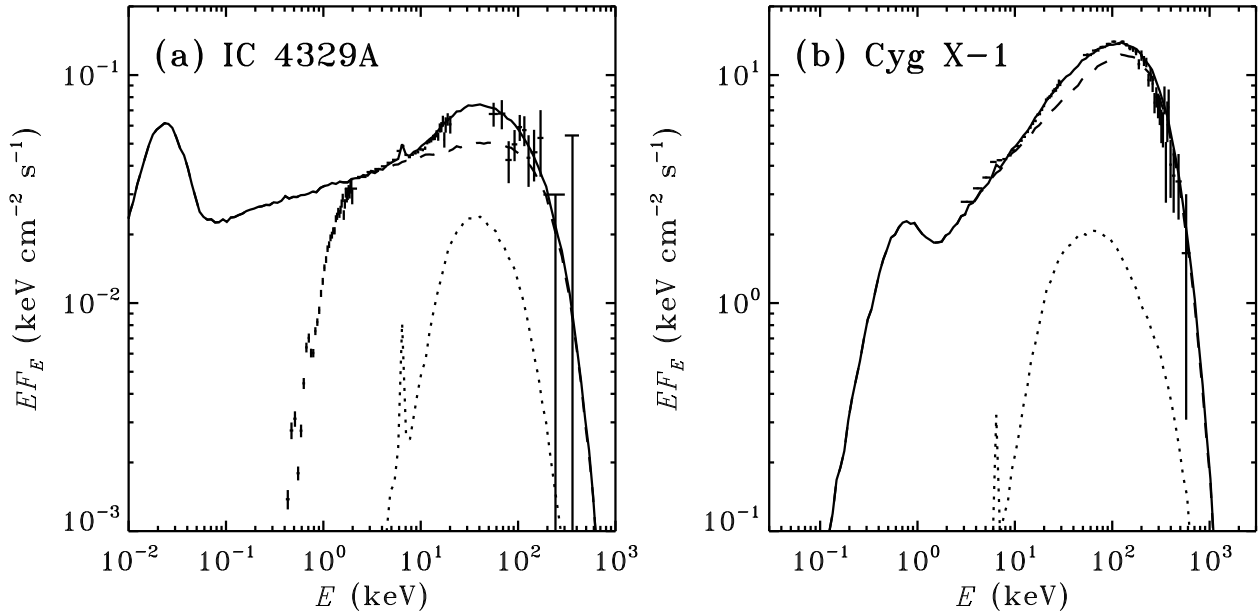,width=17cm,height=8cm}}
\caption{(a) 
Spectrum of the Seyfert 1 galaxy  IC~4329A  observed  by {\it ROSAT,  Ginga}, 
and {\it CGRO}/OSSE  (crosses,  data  from  Madejski  et al.  1995).  
The low  energy
cut-off is due to the  Galactic  absorption.  The solid  curve shows the model
spectrum  for  $\tT=3$,  $h/r=2$,  $\beta=0.1$  at  inclination   $i=40\degr$.
(b) Spectrum of Cyg~X-1 as observed by {\it Ginga} and  {\it CGRO}/OSSE 
in 1991 September  (crosses, set 2 from  Gierli\'nski  et al.  1997).  
The solid curve shows
the model spectrum for $\tT=3$, $h/r=1.25$, $\beta=0.3$ at inclination
$i=50\degr$.  In both  panels,  the dotted  curves  give  the  reflected
components and the dashed curves show the intrinisic Comptonized spectra.
}\label{fig:datasp}
\end{figure*}

Fig.~\ref{fig:datasp}  shows the spectra of one GBH source,  Cyg~X-1, 
and one bright Seyfert 1 galaxy, IC~4329A, together with model spectra.
We did not use a real fitting  procedure (since each simulation takes a few
hours on a modern  workstation),  instead,  we used  unfolded  spectra  and
searched  for similar  spectra in our sample of models.  
We find that the spectrum of Cyg~X-1 is well reproduced by the model with
a bulk  velocity  $\beta=0.3$, an optical  depth $\tT=3$, and $h/r=1.25$.
The  inclination  that  gives  the best  agreement  with  the  data,
$i=50\degr$, is compatible with the current estimates  $25\degr \leq i \leq
67\degr$ (see Gierli\'nski et al. 1999 and references therein).

For IC~4329A, we find $\beta=0.1$,  $\tT=3$, and $h/r=2$.  The inclination,
$i=40\degr$, is consistent with the Seyfert unification scheme stating that
Seyfert  1s  should  be  seen  rather  face-on   (Antonucci  1993).  
Note that $h/r$ is smaller in the case of Cyg~X-1.

\subsection{$R-\Gamma$ correlation}

\begin{figure*}
\centerline{\epsfig{file=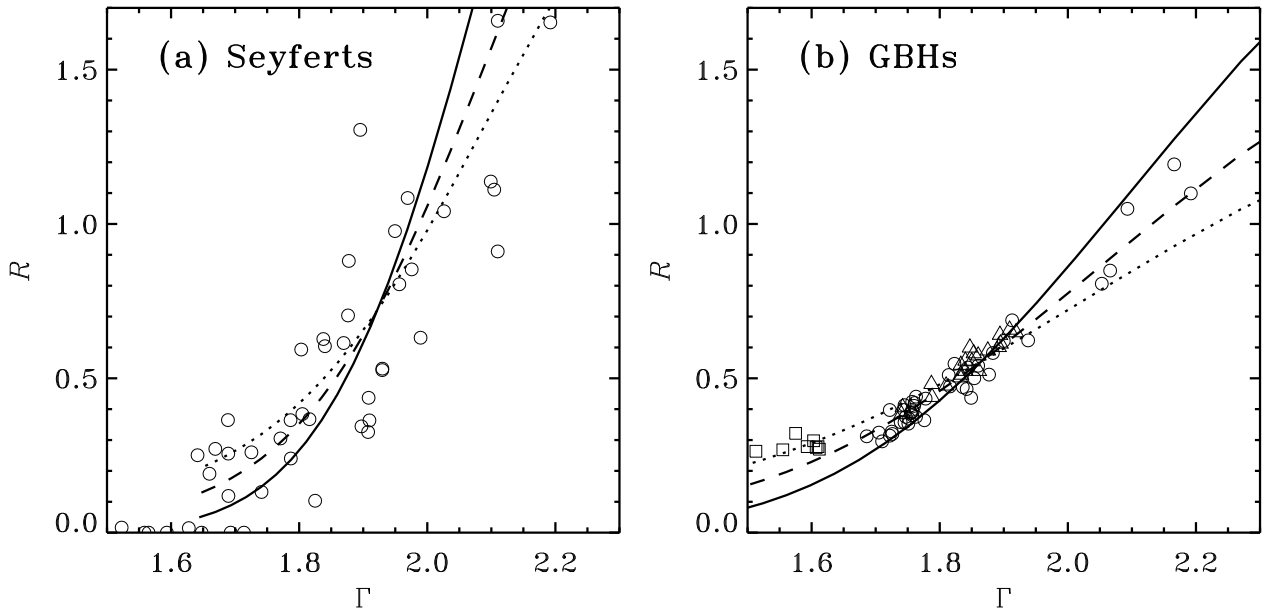,width=17cm,height=8cm}}
\caption{(a) 
The  observed  $R-\Gamma$  correlation  for  Seyfert  galaxies
(circles).  Data  from   Zdziarski  et  al.  (1999).  The  curves  show  the
correlation     predicted    by    the    model    (eqs.  \ref{eq:mus},
\ref{eq:A}--\ref{eq:reflbulk}).    Here    $\mus=\mur=0.6$,    $\tT=3$,
$a=0.15$.  The solid,  dashed, and dotted curves  correspond  to $\mu=0.85,
0.55$,   and   0.35   (i.e.   $i=32\degr$,   $57\degr$,   and   $70\degr$),
respectively.
(b) The observed  $R-\Gamma$  correlation  for GBHs, Cyg~X-1  (circles), GX
339-4  (triangles),  and GS 1354-644  (squares).  Data from Gilfanov et al.
(2000).  The curves  correspond  to the same three  inclinations  as on the
left panel.  Here $\mus=\mur=0.45$, $\tT=3$, $a=0.15$.
}\label{fig:data}
\end{figure*}

Fig.~\ref{fig:data}  gives the observed $R-\Gamma$ correlation for Seyferts
and  GBHs (Zdziarski et al.  1999; Gilfanov et al.  1999). The correlation
exists in individual objects observed at different periods as well as in a 
sample of sources.  Such a  correlation  supports  the  hypothesis  that  
it is the reprocessed/reflected radiation that cools the X-ray source via 
Comptonization.

The $R-\Gamma$ correlation expected in the dynamic corona model is similar
in shape to the observed one, and we now try to fit the data with the model.
As we showed in Section~4, the analytical model is a sufficiently good 
approximation to the results of exact simulations and it is reasonable to 
use this simple model in a fitting procedure instead of the exact transfer
model (especially taking into account the uncertainties in the validity of
assumptions such as homogeneous density and uniform heating of the blob).

The analytical model yields explicit expressions for $R(\beta)$ 
(eq.~\ref{eq:reflbulk}) and
$\Gamma(\beta)$ (eqs.~\ref{eq:A} and \ref{eq:gam_ampl}) for given parameters
$\mus$, $\mur$, $a$, and $\tT$. For simplicity, we assume $\mus=\mur$ and
fix the albedo at a reasonable value $a=0.15$. The fit to  $R-\Gamma$ 
correlation is shown in Fig.~\ref{fig:data} for AGN and GBHs.

It is interesting to note that the correlations for AGN and GBHs cannot be
fitted with the same value of $\mus$.  We find $\mus$  being lower
for GBHs  ($\mus=0.45$  or  equivalently  $h/r=1$)  than for AGN
($\mus=0.6$ or $h/r=1.5$).  This might indicate different
geometries of emission  regions in GBHs and AGN.

\subsection{The beaming effects}

The observed flux $\Fh$ of the hard X-rays beamed away from the disc can 
dominate over the blackbody component $\Fbb$ when the disc is observed
at small inclinations. The hard X-rays can dominate even if only a modest 
fraction $f$ of the accretion energy is dissipated in the corona. 
In the case of a patchy corona, the model predicts 
\be
\frac{\Fbb}{\Fh}\approx R(i)2\cos i 
\left(   1+ \frac{1-f}{f} \ \frac{1+d}{d} \right).
\ee
Here $d=L_-/L_+$ is the ratio of the Comptonized luminosities 
emitted towards and away from the disc,
\be
d=\frac{1+\beta/2}{1-\beta/2} \ \frac{(1-\beta)^2}{(1+\beta)^2} .
\ee 
For example, in the case of Cyg~X-1, taking $i=50\degr$ and $\beta=0.3$,
one gets $\Fbb/\Fh\approx 0.4$ for $f=3/4$ or 
$\Fbb/\Fh\approx 0.85$ 
for $f=1/2$,  in agreement with the data (e.g. Di Salvo et al. 2000).

The beaming of the coronal X-rays may be tested in a sample of hard-state 
objects with known inclinations: $R$ should increase with $i$. Also, the 
shape of $R-\Gamma$ correlation varies with $i$ (Fig.~\ref{fig:rvsmu}), which 
may be checked in 
the data on GBHs. Unfortunately, in the case of AGN, the high inclination 
objects are obscured by the large-scale molecular torus which dominates
the reflection component.

The ejection model with $\beta<0$ predicts very soft spectra and large ratios
$\Fbb/\Fh$. 
This situation might take place in some soft objects, e.g. Narrow Line 
Seyfert~1 galaxies (see Brandt 1999 for a review).


\section{Discussion}
\label{sec:discuss}

\subsection{Variability}

The coronal emission should be variable, with a temporal pattern governed by 
the complicated processes of the magnetic field buoyancy and its dissipation 
in the corona. The ejection velocity can vary from flare to flare and 
the fit parameters of the observed spectrum should be taken with caution
since they give only effective/average values. 

The main X-ray emitting region probably extends through a decade in radius, 
and $\beta$ is likely to be different at different radii. Recent temporal 
studies (e.g. Gilfanov et al.  2000; Revnivtsev et al. 2000) suggest that 
harder spectra with lower reflection come from the inner radii. 
In the context of the dynamic corona
model it would correspond to an increase of $\beta$ in the innermost region.
The analysis of Cyg~X-1 and GX~339-4 at different periods of the hard state
showed that the typical variability time-scales grow when the spectrum 
hardens. This behaviour may indicate that the accretion is slower in the 
hardest periods (Beloborodov 2001).

\subsection{Pair envelopes in quasars}

In this paper, we neglected scattering on the particles escaped or created
in $\gamma-\gamma$ collisions outside the heated region.
In the brightest sources, with luminosities comparable to the Eddington
limit, the hard X-rays from the flares will interact with each other and 
create an optically thick $e^\pm$ outflow covering the whole inner region 
of the accretion disc (Beloborodov 1999c). Such an outflow is relatively
cold (at the 
Compton temperature, $k\TC\sim 1-10$~keV) and it has an equilibrium bulk 
velocity such that the radiative pressure from the disc is balanced by the 
Compton drag. The velocity is mildly relativistic ($\beta$ increases from 
0.3 at the base to 0.7 at the photosphere of the outflow) and the scattering
of the disc/corona radiation in the outflow strongly affects the
angular distribution of the observed luminosity. The reflection 
amplitude will then be suppressed whatever dynamics takes place in the 
compact flares hidden in the $e^\pm$ outflow.

\subsection{Optical and UV polarization of Seyferts and quasars}

The scattering of soft radiation in a dynamic corona is accompanied by 
a strong effect that can be tested in optical/UV observations:
the polarization of scattered radiation is sensitive to $\beta$.
The polarization is perpendicular to the disc 
if $0.12<\beta<0.78$ and parallel otherwise (Beloborodov 1998).
The observed optical  polarization in non-blazar AGN is 
parallel to the radio jet that is  presumably  perpendicular  to the
accretion disc (Stockman,  Moore \& Angel 1984); it is consistent with 
the scattering in a mildly relativistic outflow.

In three of ten quasars with measured UV polarization, a
steep rise in polarization was observed blueward of the Lyman limit 
(see Koratkar  \& Blaes  1999 for a  review).  
The outflowing corona model may provide a natural explanation for this
mysterious rise (Beloborodov \& Poutanen 1999).

\section{Conclusions}

We performed  detailed  Monte-Carlo  simulations of X-ray production by
hot coronae  atop  accretion  discs, and tested the model  against the data.
The main results are as follows.

\begin{enumerate}

\item A static  corona atop a neutral  reflector is not able to produce the
observed hard   spectra with low  reflection.  Furthermore,  changes in
the coronal  geometry  produce an  anticorrelation  between  the hard X-ray
spectral  slope  $\Gamma$  and  the  amplitude  of  reflection   $R$.  This
anticorrelation is opposite to what is observed.

\item The  disc-corona  model becomes  consistent  with the data if the hot
plasma is dynamic, i.e.  moves with a mildly  relativistic  bulk  velocity,
$\beta=v/c$,  with  respect  to the  accretion  disc.  In  particular,  the
spectrum of Cyg~X-1 is reproduced by the model with $\beta=0.3$, confirming
the estimate of B99b.

\item The observed  $R-\Gamma$  correlation  is well  explained  by varying
$\beta$.  It suggests  that $\beta$ may be the main  parameter  controlling
the X-ray spectrum.

\item  The  results  of the  simulations  are in good  agreement  with  the
analytical  description  of B99a,b.  We improved  the  analytical  model by
accounting  for the  attenuation  of the  reflection  component  by the hot
plasma atop the disc (eq.~\ref{eq:reflbulk}).

\item The optical and UV  polarization  properties of AGN are sensitive
to  the  dynamics  of  the  corona.  The   polarization   data  provide  an
independent indication for the mildly relativistic bulk motion.

\end{enumerate}

\section*{ACKNOWLEDGMENTS}

This work was supported by the Italian MURST grant  COFIN98-02-15-41  (JM),
the Swedish  Natural  Science  Research  Council (AMB, JP), the  Anna-Greta  
and Holger  Crafoord  Fund (JP), and RFBR grant 00-02-16135  (AMB).  
We thank  Andrzej  Zdziarski  and Marat  Gilfanov  for providing the 
$R-\Gamma$ correlation data, and Andrei F. Illarionov for comments on the 
manuscript.


\begin{thebibliography}{}

\bibitem[]{and82}
Anders E., Ebihara M., 1982, Geochim.\ Cosmochim.\ Acta, 46, 2363

\bibitem[]{ant93}
Antonucci R., 1993, ARA\&A, 31, 473

\bibitem[ ]{bel1}
Beloborodov A. M., 1998, ApJ, 496, L105

\bibitem[ ]{bel4}
Beloborodov A. M., 1999a, in Poutanen J., Svensson R., eds,
ASP Conf. Series Vol. 161, High Energy Processes in Accreting Black Holes.
Astron. Soc. Pac., San Francisco, p. 295 (B99a)

\bibitem[ ]{bel2}
Beloborodov A. M., 1999b, ApJ, 510, L123 (B99b)


\bibitem[ ]{bel3}
Beloborodov A. M., 1999c, MNRAS, 305, 181

\bibitem[ ]{bel5}
Beloborodov A. M., 2001, Adv. Space Res., in press

\bibitem[ ]{bp99}
Beloborodov A. M., Poutanen J., 1999, ApJ, 517, L77

\bibitem[ ]{bkb77}
Bisnovatyi-Kogan G. S., Blinnikov S. I., 1977, A\&A, 59, 111

\bibitem[ ]{bra99}
Brandt W. N., 1999, in Poutanen J., Svensson R., eds,
ASP Conf. Series Vol. 161, High Energy Processes in Accreting Black Holes.
Astron. Soc. Pac., San Francisco, p. 166

\bibitem[ ]{dis00}
Di Salvo T., Done C., \.Zycki P. T., Burderi L., Robba N. R., 2000,
ApJ, in press (astro-ph/0010062)

\bibitem[ ]{esin}
Esin A. A., Narayan R., Cui W., Grove J. E., Zhang S.-N., 1998, ApJ, 505, 854

\bibitem[ ]{grv}
Galeev A. A., Rosner R., Vaiana G. S., 1979, ApJ, 229, 318

\bibitem[ ]{gf91}
George I. M., Fabian A. C., 1991, MNRAS,  249,  352

\bibitem[ ]{gier1}
Gierli\'nski M., Zdziarski A. A., Done C., Johnson W. N., Ebisawa K., Ueda Y.,
Haardt F., Phlips B. F., 1997, MNRAS, 288, 958

\bibitem[ ]{gier2}
Gierli\'nski M., Zdziarski A. A., Poutanen J.,  Coppi P. S.,
Ebisawa K., Johnson W. N., 1999, MNRAS, 309, 496

\bibitem[ ]{gilf1}
Gilfanov M., Churazov E., Revnivtsev M., 1999, A\&A, 352, 182

\bibitem[ ]{gilf2}
Gilfanov M., Churazov E., Revnivtsev M., 2000,
in Proc. 5th CAS/MPG Workshop on High Energy Astrophysics, in press
(astro-ph/0002415)

\bibitem[ ]{hm93}
Haardt F., Maraschi L., 1993, ApJ, 413, 507  (HM93)

\bibitem[ ]{hmg94}
Haardt F., Maraschi L., Ghisellini G., 1994, ApJ, 432, L95

\bibitem[ ]{kb99}
Koratkar A., Blaes O., 1999, PASP, 111, 1

\bibitem[ ]{liang79}
Liang E.~P.~T., 1979, ApJ, 231, L111

\bibitem[ ]{mad95}
Madejski G. M. et al., 1995, ApJ, 438, 672

\bibitem[ ]{mz95}
Magdziarz P., Zdziarski A. A., 1995, MNRAS, 273, 837

\bibitem[ ]{mj00}
Malzac J., Jourdain E., 2000, A\&A, 359, 843

\bibitem[ ]{ms00}
Miller K. A., Stone J. M., 2000, ApJ, 534, 398

\bibitem[ ]{nkk00}
Nayakshin S., Kazanas D., Kallman T. R., 2000, ApJ, 537, 833

\bibitem[ ]{pou98}
Poutanen J., 1998, in Abramowicz M. A.,  Bj\"ornsson G., Pringle J., eds,
Theory of Black Hole Accretion Disks. Cambridge Univ. Press, Cambridge, p. 100

\bibitem[ ]{ps96}
Poutanen J., Svensson R., 1996, ApJ, 470, 249 (PS96)

\bibitem[ ]{pkr97}
Poutanen J., Krolik J. H., Ryde F., 1997, MNRAS, 292, L21

\bibitem[ ]{rev1} 
Revnivtsev M., Gilfanov  M., Churazov E., 1999, A\&A, 347, L23
 
\bibitem[ ]{rev2} 
Revnivtsev M., Gilfanov  M., Churazov E., 2000, A\&A, submitted
(astro-ph/9910423)

\bibitem[ ]{rfy99}
Ross R. R., Fabian A. C., Young A. J., 1999, MNRAS, 306, 461

\bibitem[ ]{rl79}
Rybicki G. B., Lightman A. P., 1979, Radiative Processes in Astrophysics.
Wiley, New York

\bibitem[ ]{st85}
Stern B. E., 1985, SvA, 29, 306

\bibitem[ ]{st95a}
Stern B. E., Begelman M. C., Sikora M., Svensson R., 1995a, MNRAS, 272, 291

\bibitem[ ]{st95b}
Stern B. E., Poutanen J., Svensson R., Sikora M., Begelman M. C., 1995b,
ApJ, 449, L13

\bibitem[ ]{sto84}
Stockman H. S., Moore R. L.,  Angel J. R. P., 1984, ApJ, 279, 485

\bibitem[ ]{sve96}
Svensson R., 1996, A\&AS, 120, 475

\bibitem[ ]{tp92}
Tout C. A., Pringle J. E., 1992, MNRAS, 259, 604

\bibitem[ ]{war00}
Wardzi\'nski G.,  Zdziarski A. A., 2000, MNRAS, 314, 183

\bibitem[ ]{zdz97}
Zdziarski A. A., Johnson W. N., Poutanen J., Magdziarz P., Gierli\'nski M.,
1997, in Winkler C., Courvoisier T. J.-L., Durouchoux Ph., eds,
Proc. 2nd INTEGRAL Workshop, The Transparent Universe, ESA SP-382.
  ESA, Noordwijk.  p. 373

\bibitem[ ]{zls99}
Zdziarski A. A., Lubi\'nski P., Smith D. A., 1999, MNRAS, 303, L11

\bibitem[ ]{zyc94}
\.Zycki P. T., Krolik J. H., Zdziarski A. A., Kallman T. R., 1994, ApJ, 437, 597
\end{thebibliography}
\end{document}